\documentstyle[epsfig]{europhys}

\newif\ifboo \boofalse

\begin{document}
\euro{53}{2}{251}{2001}
\Date{}
\shorttitle{Mahn-Soo Choi {\it et al.} Coherent oscillations}
\title{Coherent oscillations in a Cooper-pair box}
\author{Mahn-Soo Choi\inst{1,4}, Rosario Fazio\inst{2,3}, 
	Jens Siewert\inst{2,3}, and C. Bruder\inst{1}}
\institute{
   \inst{1} Departement Physik und Astronomie, Universit\"at Basel,
Klingelbergstrasse 82, CH-4056 Basel, Switzerland\\
\inst{2}Dipartimento di Metodologie Fisiche e Chimiche (DMFCI),
Universit\`a di Catania, viale A. Doria 6, I-95125 Catania,
Italy\\
 \inst{3}Istituto Nazionale per la Fisica della Materia (INFM),
Unit\`a di Catania, Italy\\
\inst{4} Korea Institute for Advanced Study, Cheongryangri-dong 207-43,
  Seoul 130-012, South Korea}

\rec{}{}
\maketitle

\begin{abstract}
This paper is devoted to an analysis of the experiment by Nakamura
{\it et al.} (Nature {\bf 398}, 786 (1999)) on the quantum state
control in Josephson junctions devices. By considering the relevant
processes involved in the detection of the charge state of the box and
a realistic description of the gate pulse we are able to analyze some
aspects of the experiment (like the amplitude of the measurement
current) in a quantitative way.
\end{abstract}

The possibility to form coherent superpositions of states is one of
the most fundamental properties that distinguishes quantum from
classical physics. On the microscopic level, many examples come to
mind. Whether it is possible, however, to superpose macroscopically
distinct quantum states is controversial and has been debated since
the advent of quantum mechanics. Whereas at the beginning it was
thought that the macroscopic world is in some sense classical, there
have been a number of suggestions over the years on how it might be
possible to observe macroscopic quantum coherence in solid-state
devices~\cite{leggett}. Superconducting nanocircuits have been used
successfully to test quantum mechanics in mesoscopic systems.
Examples are the test of the Heisenberg uncertainty principle in a
mesoscopic superconductor~\cite{matters} or experiments on the
superposition of charge states in Josephson
junctions~\cite{Matters95,Bouchiat98}. More recently, the increasing
interest in quantum computation~\cite{ekert} and the search for
implementations that can be scaled and integrated has made
superconducting circuits promising candidates to realize
qubits~\cite{qubits}.

In a recent experimental breakthrough, Nakamura {\it et
al.}~\cite{Nakamu99} demonstrated coherent oscillations between two
charge states of a superconducting island in a single-electron device,
the so-called Cooper-pair box. These states are macroscopically
distinct in the sense that they are two different states of an island
that contains, say, $10^{8}$ electrons. The superposition is achieved
by switching the gate voltage of the box quasi-instantaneously to the
point at which the two charge states, $n=0$ and $n=2$, are
degenerate. Here $n$ is the number of excess charges on the island
(one excess Cooper pair means $n=2$). As a result, the system performs
Rabi oscillations between these two states that are monitored by
another weakly coupled tunnel junction.

In this paper, we analyze the experiment by Nakamura
{\it et al.}~\cite{Nakamu99} by solving the appropriate master
equation. The reason for performing the work presented here is
twofold. Although the description of the coherent oscillations in
terms of two-state system is appealing and contains most of the
physics there are fundamental questions which are still
unanswered. They are related to the mechanisms of decoherence and,
particularly important for this experiment, the role of the measuring
apparatus and real shape of the gate pulse. Each of these issues leads
to computational errors when the Cooper-pair box is used as a
(charge)-qubit. Therefore to investigate each of these issues in great
detail is a necessary prerequisite for the implementation of a solid
state quantum computer with Josephson nanocircuits. The next step in
the coherent control of the dynamics of a macroscopic system is the
experimental verification of conditional dynamics, i.e. of
entanglement. Entangled states, a central concept of quantum
mechanics, are difficult to characterize and measure. A quantitative
understanding of the single qubit experiment is also relevant for
modeling two-qubit gates.

\begin{figure}\centering
\epsfig{file=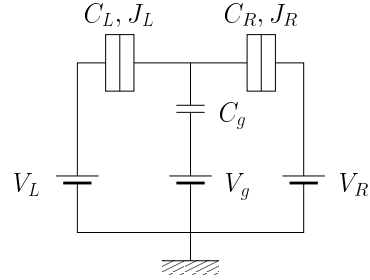,width=60mm}
\caption{Equivalent circuit. The Cooper-pair box corresponds to the
configuration $J_L=E_J$, $J_R=0$, $eV_L=0$ and $eV_R=-eV\simeq{2\Delta}$
using the right lead as a probe gate.}
\label{cpbox:fig1}
\end{figure}

To investigate the results obtained in Ref.~\cite{Nakamu99}, we consider 
the superconducting transistor shown in Fig.~\ref{cpbox:fig1} and 
described by the Hamiltonian (see, e.g., \cite{Ingold92})
\begin{equation} \label{cpbox:1}
H = \frac{(Q+Q_t)^2}{2C} - \sum_{j=L,R}Q_jV_j
- \sum_{j}J_j\cos\phi_j \;.
\end{equation}
The first two terms define the charging part ($H_C$), for which we
have adopted the effective capacitance model: $C=C_L+C_R+C_g$ with
junction capacitance $C_L$ and $C_R$ for the left and right junction
and $C_g$ for the gate. When the leads and the gate are biased by
voltages $V_L$, $V_R$, and $V_g$, respectively, the total gate-induced
charge is given by $Q_t=C_LV_L+C_RV_R+C_gV_g$. $Q_j/e$ is the number
of electrons which have passed through the junction $j$ to the
central electrode D, and $Q=Q_L+Q_R$ is the total charge on D. The
last term in (\ref{cpbox:1}) is the Josephson part ($H_J$) with the
Josephson coupling energy $J_j$ and the phase difference $\phi_j$
across the junctions $j=L,R$. In the Coulomb-blockade regime, where
the charging energy $E_C=e^2/2C$ is larger than $J_j$,
the charge $Q$ on the island is quantized in units of electric charge
$e$. Accordingly, we will use the basis of charge states $|n\rangle$
with charge $Q=ne$ where $n$ is integer.

The coupling of the system described by Eq.~(\ref{cpbox:1}) to the
environment leads to decoherence. The environment or bath that we
focus on here are the quasiparticles on the two superconducting leads
$L,R$ and the central island $D$
\begin{equation} \label{cpbox:2}
H_{{qp}}
= \sum_{\alpha=L,R,D}\sum_{k\sigma} 
\varepsilon_{k\alpha}\gamma_{k\alpha\sigma}
^\dag\gamma_{k\alpha\sigma}^{}\; .
\end{equation}
Here, $\gamma_{k\alpha}^\dag$ ($\gamma_{k\alpha}$) creates
(annihilates) a quasiparticle with momentum $k$ and energy
$\varepsilon_{k\alpha} = \sqrt{\xi_{k\alpha}^2+\Delta^2}$ in electrode
$\alpha$. Also, $\xi_k$ is the usual single-particle dispersion (with
respect to the chemical potential), $\Delta$ is the superconducting
gap (to simplify the discussion, we assume identical superconductors,
as bulk, for all electrodes) and $\sigma$ labels the spin. The effects
of quasiparticle tunneling can be described by the tunneling
Hamiltonian
\begin{equation} \label{cpbox:3}
H_T = \sum_{j=L,R}\left[ e^{-i\phi_j/2}\,\,X_j + h.c.\right] \;,
\end{equation}
where
$X_j = \sum_{kq\sigma}T_{kq}\gamma_{kj\sigma}^\dag \gamma_{qD\sigma}^{}$
and $T_{kq}$ is the tunneling amplitude. 
The total Hamiltonian is given by $H_{{tot}}=H+H_{{qp}}+H_T$.

The evolution in time of the reduced density matrix
$\rho={{Tr}}_{\gamma^\dag,\gamma}\rho_{{total}}$
is governed by the generalized master equation 
(see, e.g., \cite{Blumxx96})
\begin{eqnarray}
\nonumber\lefteqn{
\dot\rho(t) = - \frac{i}{\hbar}\left[H,\rho(t)\right]
}\\&& \label{cpbox:4}
- \sum_j\int_0^\infty{ds}\;\alpha_j^>(s)
 \left[ e^{-i\phi_j/2}, e^{+i\phi_j(-s)/2}\rho(t)\right]-h.c.
 \\&& \nonumber
- \sum_j\int_0^\infty{ds}\;\alpha_j^<(-s)
 \left[ e^{+i\phi_j/2}, e^{-i\phi_j(-s)/2}\rho(t)\right]-h.c. \;.
\end{eqnarray}
The correlation
functions
$\alpha_j^>(t)\equiv\left\langle{X_j(t)X_j^\dag}\right\rangle/\hbar^2$ and
$\alpha_j^<(t)\equiv\left\langle{X_j^\dag
X_j(t)}\right\rangle/\hbar^2$
describe the effects of quasiparticle tunneling.
More explicitly, the Fourier transforms of these correlation functions
can be expressed in terms of the quasiparticle current-voltage
characteristics $I_j^{{qp}}(E)$:
$\alpha_j^>(E)=\left[1+n_B(E)\right]I_j^{{qp}}(E)/e$ and
$\alpha_j^<(E)=n_B(E)I_j^{{qp}}(E)/e$, where
$n_B(E)=1/\left(e^{\beta{E}}-1\right)$ is
the thermal distribution function. 

The contribution to the transport due to resonant Cooper-pair
tunneling has been considered by Averin and Aleshkin~\cite{Averin89}
and by van den Brink {\it et al.}~\cite{Brinkx91a}. They were
interested in the d.c. transport current. In the present paper, we
focus on the coherent oscillation between the charge states and its
decoherence due to quasiparticle tunneling. In this case, the
off-diagonal elements play a crucial role in the dynamics of the
reduced density matrix. We note in passing that it is also possible to
use $H_C$ instead of $H$ as unperturbed Hamiltonian~\cite{Averin89}
and the Josephson tunneling $H_J$ in (\ref{cpbox:1}) and the
quasiparticle tunneling $H_T$ in (\ref{cpbox:3}) as perturbations;
this is equivalent to our model to second order in the tunneling
amplitude.

The Cooper-pair box is probed via quasiparticle
tunneling~\cite{Nakamu99}.  This can be achieved by configuring the
system parameters as follows: the whole system is biased so that 
Cooper-pair tunneling occurs only across the left junction whereas there
is only quasiparticle tunneling across the right junction, $eV_L=0$ and
$eV_R=-eV\sim-2\Delta$. Respectively, we can safely set $J_L=E_J$ and
$J_R=0$, and it is always implied that
$k_BT\ll{}E_J\ll{}E_C\ll\Delta$.  Due to the strong Coulomb repulsion
$E_J\ll{}E_C$, it suffices to consider the subspace of $\{|0\rangle,
|1\rangle, |2\rangle\}$. In this basis, the time evolution of the
diagonal elements $\sigma_n\equiv\rho_{n,n}$ ($n=0,1,2$) and the
off-diagonal element $\chi\equiv\rho_{0,2}$ is given
by~\cite{Averin89}
\begin{eqnarray}\label{cpbox:5}
\label{cpbox:5a}
\dot\sigma_0(t)
& = & - i\frac{E_J}{2\hbar}\left[ \chi - \chi^* \right] 
 - \Gamma^+(0)\sigma_0(t) + \Gamma^-(1)\sigma_1(t) \\
\label{cpbox:5b}
\dot\sigma_2(t)
& = & + i\frac{E_J}{2\hbar}\left[ \chi - \chi^* \right]
 - \Gamma^-(2)\sigma_2(t) + \Gamma^+(1)\sigma_1(t) \\
\label{cpbox:5c}
\dot\chi(t)
& = & + i\frac{E_{2,0}(t)}{\hbar}\chi(t)
 - i\frac{E_J}{2\hbar}\left[ \sigma_0 - \sigma_2 \right]
 \\&&\mbox{}\nonumber
 - {1 \over 2}\left[\Gamma^+(0)+\Gamma^-(2)\right]\chi(t) \;,
\end{eqnarray}
together with the normalization condition 
${{Tr}}\rho(t)=\sigma_0(t)+\sigma_1(t)+\sigma_2(t)=1$. Here
$E_{2,0}=4[1-Q_t(t)/e]E_C$ is the change in the charging energy when a
Cooper pair tunnels into the Cooper-pair box across the left junction.
In Eqs.~(\ref{cpbox:5}),
$\Gamma^\pm(n)=\Gamma_L^\pm(n)+\Gamma_R^\pm(n)$ and $\Gamma_j^\pm(n)$
is the quasiparticle tunneling rate across the junction $j=L,R$
resulting in the transition $n\to{n\pm1}$. These rates can be written
explicitly as
\begin{eqnarray}
\Gamma^+(0)
& = & \alpha_L^>(E_-) + \alpha_R^<(eV-E_-) \\
\Gamma^-(2)
& = & \alpha_L^>(E_+) + \alpha_R^>(eV+E_+) \\
\Gamma^+(1)
& = & \alpha_L^<(E_+) + \alpha_R^<(eV+E_+) \\
\Gamma^-(1)
& = & \alpha_L^<(E_-) + \alpha_R^>(eV-E_-) \;,
\end{eqnarray}
where $E_\pm=E_C\left[1\pm2(1-Q_t/e)\right]$.
It is useful to notice that at sufficiently low temperatures
($k_BT\ll\Delta$), $\Gamma^+(0)$ and $\Gamma^+(1)$ are exponentially small
($\sim{}e^{-\Delta/k_BT}$) while $\Gamma^-(2)\simeq\Gamma^-(1)$.
From (\ref{cpbox:5}), it is clear that quasiparticle tunneling causes
decoherence in the system. Indeed, at $Q_t=e$, the coherent oscillation
between the two degenerate charge states $|0\rangle$ and $|2\rangle$ decays
with the time scale
$1/\Gamma\equiv{}1/\Gamma^-(2)\simeq{}1/\Gamma^-(1)$,
\begin{eqnarray}
\sigma_0(t)
& \simeq & \frac{1}{3}+\frac{e^{-3\Gamma{t}/2}}{6}
 + \frac{e^{-\Gamma{t}/2}}{2}\left(\cos\omega{t}+\frac{\hbar\Gamma}{E_J}
\sin\omega{t}\right)
 \\
\sigma_2(t)
& \simeq & \frac{1}{3}+\frac{e^{-3\Gamma{t}/2}}{6}
 - \frac{e^{-\Gamma{t}/2}}{2}\cos\omega{t}
 \\
i\chi(t)
& \simeq & \frac{\hbar\Gamma}{3E_J}
 - \frac{\hbar\Gamma e^{-3\Gamma{t}/2}}{12E_J}\nonumber\\
 && + \frac{e^{-\Gamma{t}/2}}{4}\left(2\sin\omega{t}-\frac{\hbar\Gamma}{E_J}
\cos\omega{t}\right) \;,
\end{eqnarray}
where the oscillation frequency is also slightly modified as
$\omega\simeq(E_J/\hbar)\left[1+\left(\hbar\Gamma/E_J\right)^2\right]$.
Here we assumed the initial conditions $\sigma_0(0)=1$ and
$\sigma_2(0)=\chi(0)=0$. In the {\em stationary limit\/}
($t\to\infty$), the charge states $|0\rangle$, $|1\rangle$, and
$|2\rangle$ are equally populated. This corresponds to the
Josephson-quasiparticle cycle~\cite{Averin89,Brinkx91a,Fulton89}, i.e.,
resonant tunneling of Cooper pairs followed by sequential tunneling of
quasiparticles to the probe gate.

We now look more closely at the experiment by Nakamura {\it et
al.}~\cite{Nakamu99}. A pulse of finite length $\Delta{t}$ was applied
to the gate to change the total gate-induced charge from $Q_t=Q_0$ to
$Q_t=Q_0+Q_p$ and back. For example, if $Q_0/e<1$ (far from the
resonance point) and $(Q_0+Q_p)/e=1$ (at the resonance point), the
pulse causes the system (which is initially in state $|0\rangle$) to
oscillate between $|0\rangle$ and $|2\rangle$. Depending on the pulse
length $\Delta{t}$, the system may or may not be in $|2\rangle$ with
an increased probability at the end of the pulse. The decay of
$|2\rangle$ to $|0\rangle$ through quasiparticle tunneling leads to an
excess current. This was illustrated in \cite{Nakamu99} by solving the
time-dependent Schr\"odinger equation for an isolated two-level
system. Here we solve the master equation (\ref{cpbox:5}), which
includes the decoherence due to the measuring device, and calculate the
time-averaged current $I=\langle{I(t)}\rangle
=\Gamma^-(2)\langle\sigma_2(t)\rangle+\Gamma^-(1)\langle\sigma_1(t)\rangle$.
To simulate the experimental situation, we generated an array of
pulses with repetition time $T_r$ (after a few times of repetition,
the time series of $\rho(t)$ repeats the same pattern). We took the
parameters used in the experiment, i.e., $E_J = 51.8 \mu$eV, $E_C =
117\mu$eV, and $\Delta = 230\mu$eV. The voltage was chosen to be $eV
\simeq 2\Delta + 1.65E_C$, and the repetition time $T_r = 16$ns.
Figure~\ref{cpbox:2} shows the result of our calculation, viz., the
pulse-induced excess current versus offset gate charge $Q_0/e$ and/or
pulse length $\Delta{t}$.

Apart from decoherence by quasiparticle tunneling, the real
experimental situation includes several complications: (i)
the repetition time $T_r$ is finite and comparable to $1/\Gamma^-(2)$ and
$1/\Gamma^-(1)$; (ii) `jittering' of the pulse; and (iii) finite
rising/falling time of the pulse. The effect of these complications 
is summarized in Fig.~\ref{cpbox:fig3} and they are discussed below.

Ideally, the time between two pulses should be long enough (compared
to $1/\Gamma^-(2)$ or $1/\Gamma^-(1)$) such
that the system is guaranteed to relax to the desired initial
state. Experimentally, $T_r$ is limited by the detector sensitivity,
since the maximum value of the pulse-induced current is given by
$\Delta{I}_{{\rm max}}=2e/T_r$.
The value of $T_r$ in \cite{Nakamu99} was comparable to $1/\Gamma^-(2)$
or $1/\Gamma^-(1)$ and the waiting time, $T_w\equiv{}T_r-\Delta{t}$, was
not long enough.  First of all, this reduces the
pulse-induced current collected in the probe gate with respect to the ideal
maximum current $\Delta{I}_{{\rm max}}$.
And secondly, the finite $T_r$ also results in a wiggly behavior of
$\Delta{I}(\Delta{t})$: The charge oscillations with small amplitude and
large frequency ($\sim\sqrt{E_C^2+E_J^2}/\hbar$) have not decayed yet when
the next pulse is turned on, see the time dependence of $\sigma_2(t)$ in
Fig.~\ref{cpbox:fig3} (a) (dashed line).
A small change in $\Delta{t}$ with $T_r$ fixed leads to a
change in $T_w$, which may be comparable to $\hbar/\sqrt{E_C^2+E_J^2}$
and hence results in a rapid change in the pulse-induced current, see
Fig.~\ref{cpbox:fig3} (b).

\begin{figure}
\mbox{\epsfig{file=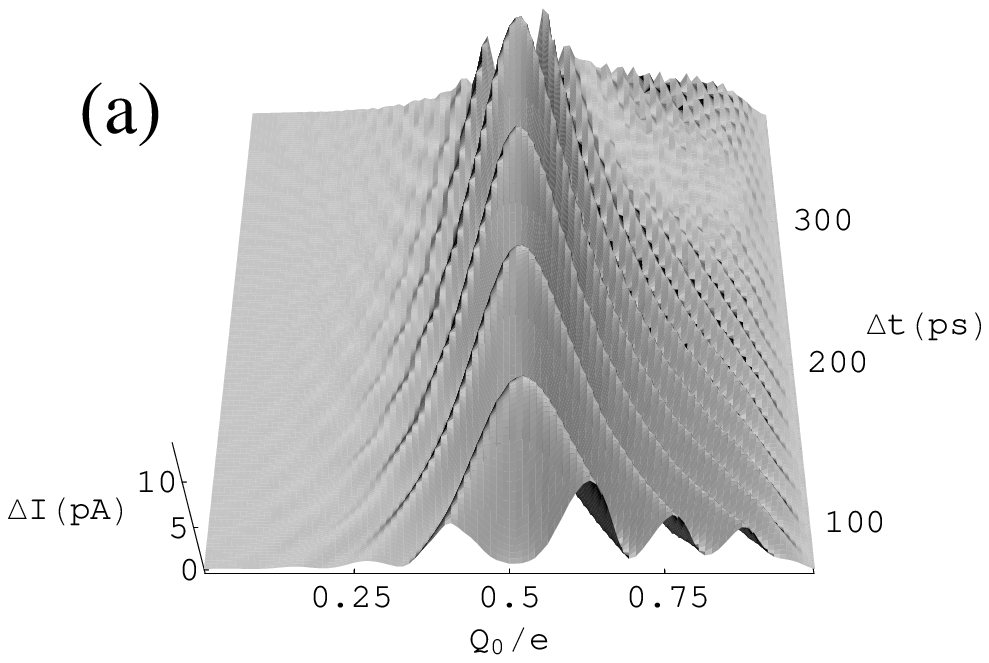,height=5.5cm}
\ \epsfig{file=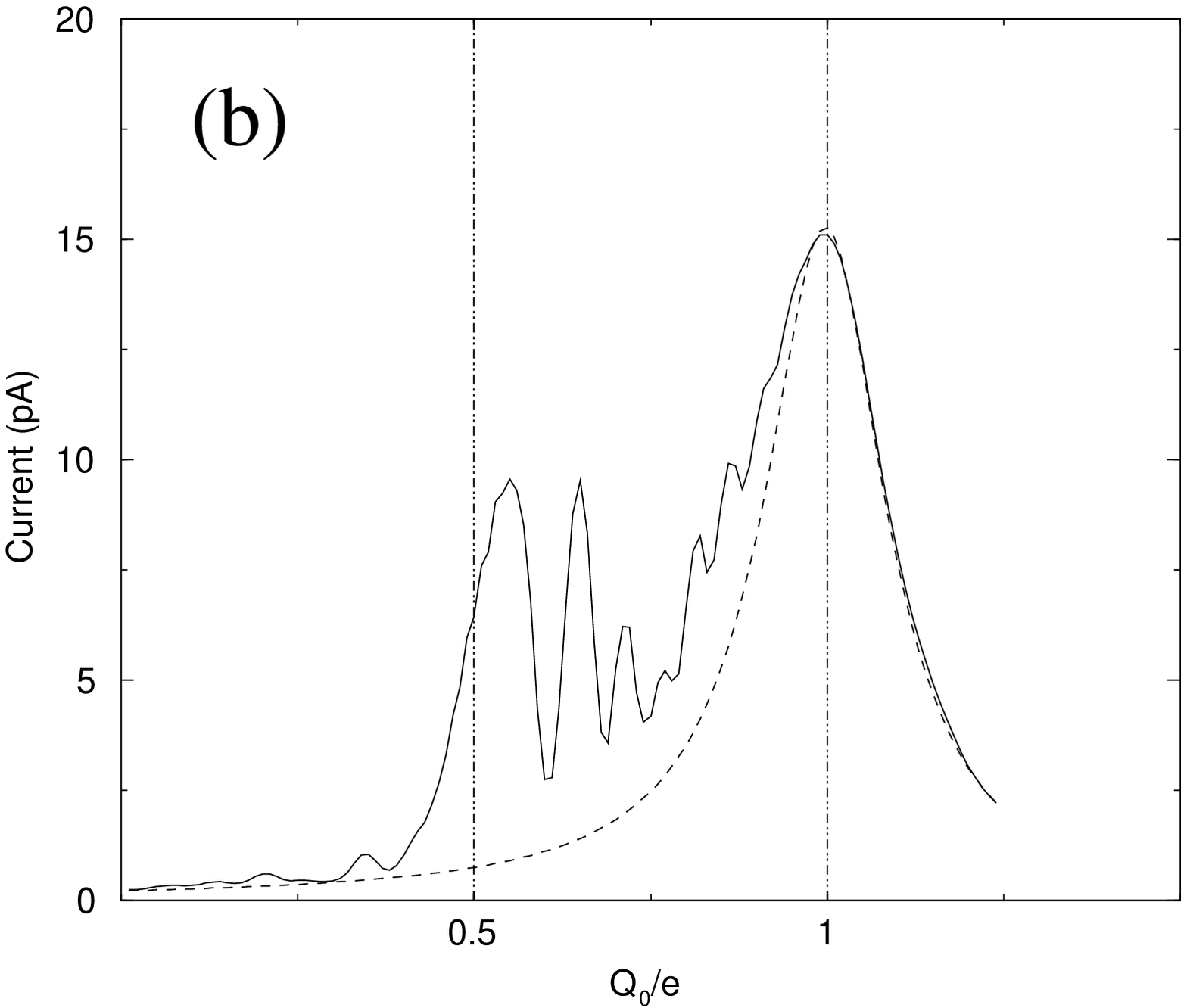,height=4.5cm}}
\caption{(a) Plot of pulse-induced current versus $Q_0/e$ and
$\Delta{t}$. (b) Current through the probe junction (R) versus
$Q_0/e$ with (solid) and without (dashed line) pulse of length
$\Delta{t}=2T_{{coh}}\simeq{}160$ps. In both plots, we used the
resistance ratio $R_L/R_R=1/1800$ which corresponds to quasiparticle
tunneling rates $\Gamma^-(2)\sim(6{\rm ns})^{-1}$ and
$\Gamma^-(1)\sim(8{\rm ns})^{-1}$, i.e., like 
in \protect\cite{Nakamu99}.}
\label{cpbox:fig2}
\end{figure}

\begin{figure}\centering
\epsfig{file=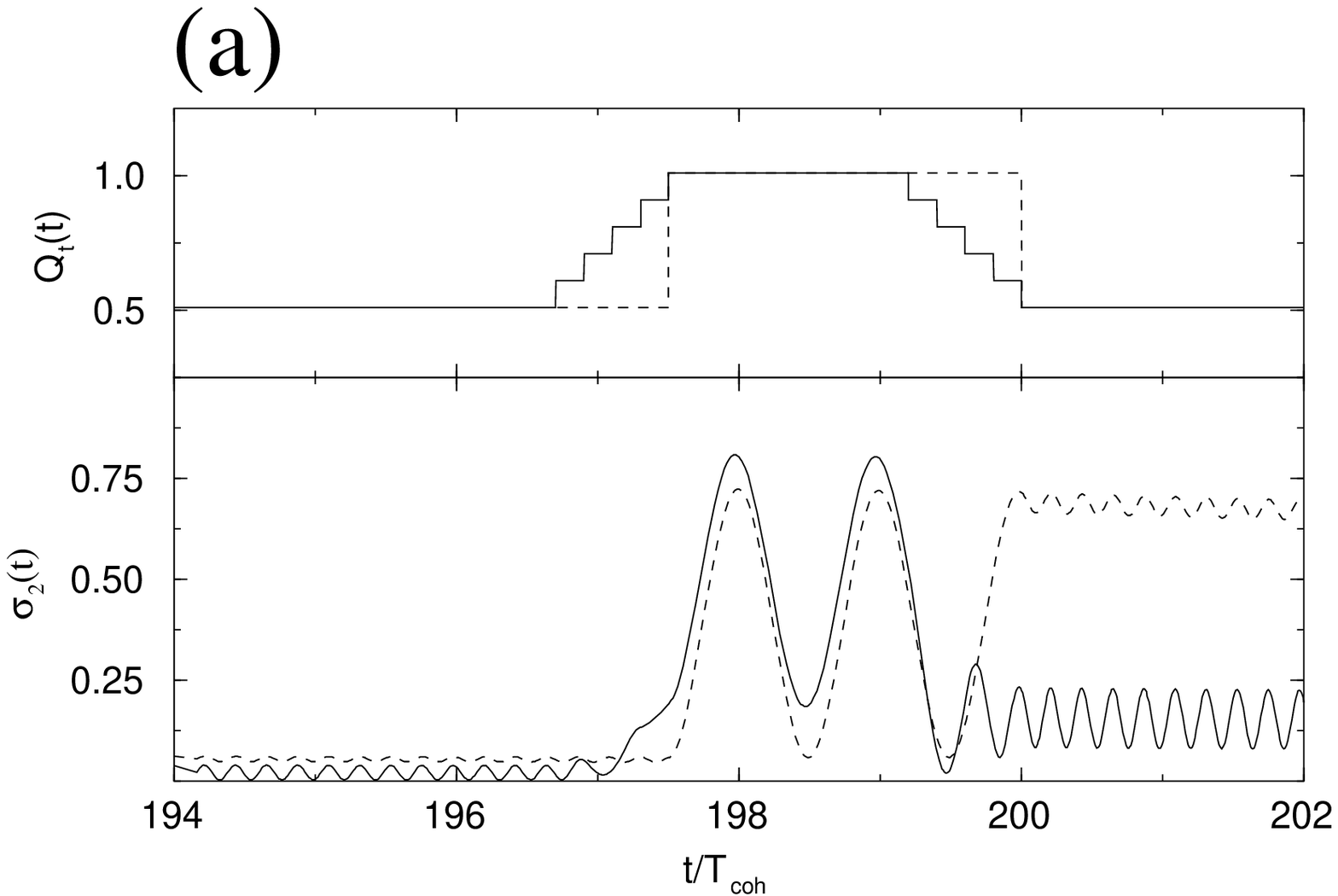,width=70mm}
\epsfig{file=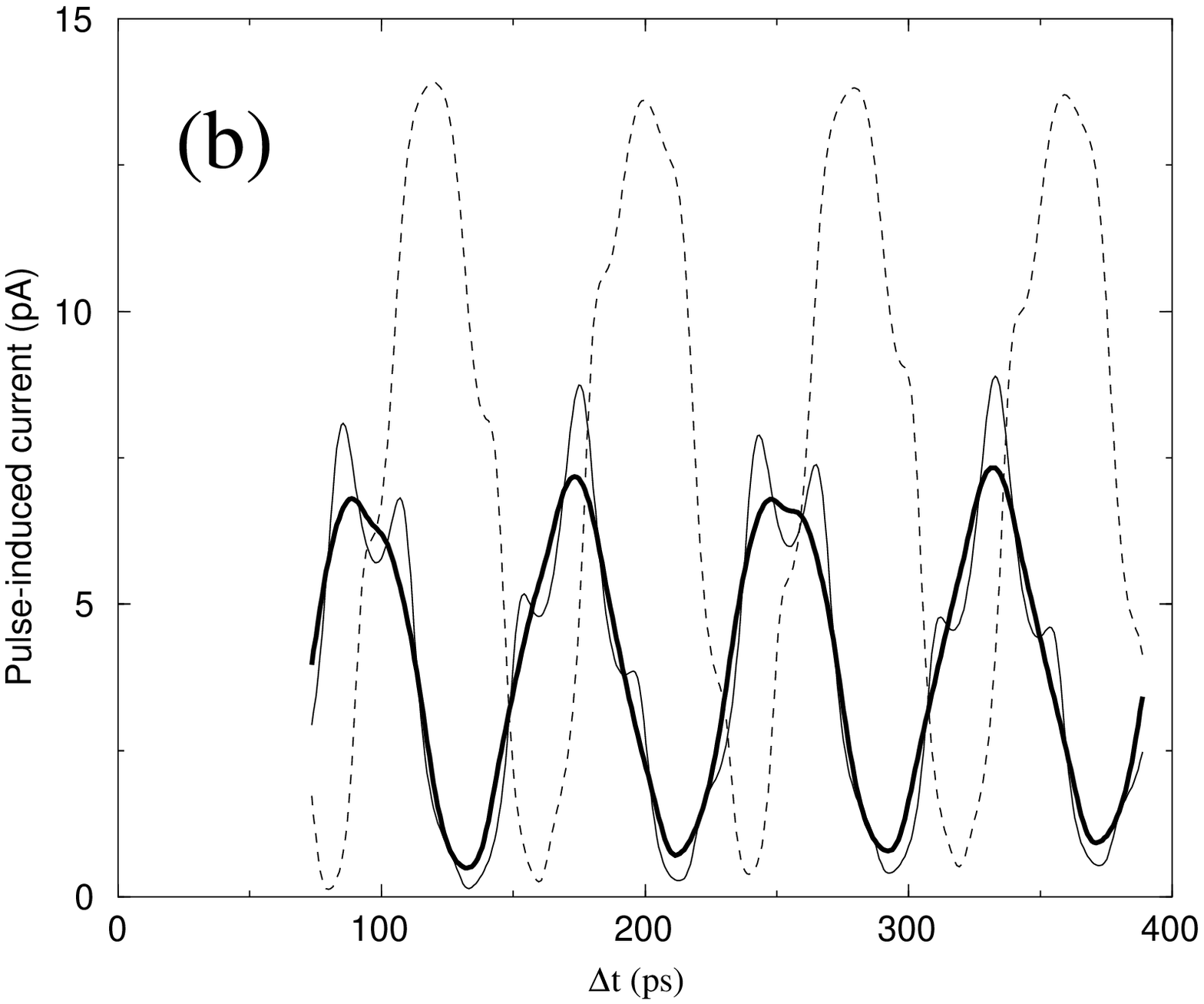,width=70mm}
\caption{(a) Pulse shapes used to create the current in (b) and
corresponding time dependence of $\sigma_2(t)$. Each step of the
step-wise pulse is $0.2T_{{coh}}$ long. The other parameters used are
$T_r=200T_{{coh}}\simeq{}16$ns, $Q_0/e=0.51$, $R_L/R_R=1/1800$
(corresponding to the values of $\Gamma^-(2)\sim(6{\rm ns})^{-1}$ and
$\Gamma^-(1)\sim(8{\rm ns})^{-1}$ in \protect\cite{Nakamu99}).  (b)
Pulse-induced current versus pulse time $\Delta{t}$ with finite (thin
solid line) and zero (dashed line) rising/falling time of the
pulse. The fat solid line shows the effect of additional averaging
by jittering of the pulse; its amplitude reproduces the experimental value.}
\label{cpbox:fig3}
\end{figure}

The jittering of the pulse leads to fluctuations in $T_r$ (or
$T_w$). This smears out the wiggly behavior in $\Delta{I}$
discussed above. The pulse jittering in \cite{Nakamu99} was of the
order of several percent of the coherence time 
$T_{\rm coh}=h/E_J$~\cite{nakamuraprivate}. 
Numerically, the jittering can be taken into account by
averaging $\Delta{I}(\Delta{t})$ over slightly different values of
$\Delta{t}$.  In Fig.~\ref{cpbox:fig2} and Fig.~\ref{cpbox:fig3} (b)
(fat solid line), we took the average over the interval
$(T_r-7{\rm ps},T_r+7{\rm ps})$.

Experimentally, the oscillation amplitude of the pulse-induced current
measured in the probe gate was significantly smaller than the ideal
maximum value $\Delta{I}_{{\rm max}}$. This may be accounted for by
taking into account the finite repetition time and finite
rising/falling time of the pulse. The effect of a finite repetition
time was already discussed above.  To understand the effect of the
finite ramping time of the pulse, we increased/decreased the gate voltage
in a step-wise way, see Fig.~\ref{cpbox:fig3} (a). As shown in
Fig.~\ref{cpbox:fig3} (b), the oscillation amplitude for this
pulse shape (thin solid line) is suppressed compared with that for an
ideal pulse (dashed line). The fat solid line shows the effect of
additional averaging by jittering of the pulse; its amplitude
reproduces the experimental value. The shift of the oscillation in
$\Delta{t}$ is due to the change in the {\em effective\/} pulse
length.

In conclusion, we have presented an analysis of the
experiment by Nakamura {\it et al.}~\cite{Nakamu99} by solving the
appropriate master equation. In particular, we have considered the
relevant processes involved in the detection of the charge state of
the box and have used a realistic description of the gate pulse.

We would like to acknowledge stimulating discussions with Y. Nakamura,
G. Falci, and E. Paladino.

\end{document}